\documentclass[%
 reprint,
superscriptaddress,
 amsmath,amssymb,
 aps,
prb,
]{revtex4-2}

\usepackage{graphicx}% Include figure files
\usepackage{dcolumn}% Align table columns on decimal point
\usepackage{bm}% bold math
\usepackage[version=3]{mhchem} % Formula subscripts using 

\begin{document}

%\preprint{APS/123-QED}

\title{ Antiferromagnetic Phases in Zr-Fe-Ge Kagome Systems}

\author{Peter Minch}
\affiliation{Department of Physics and Astronomy, Iowa State University, Ames, Iowa 50011, USA}
\affiliation{Ames Laboratory-USDOE, Iowa State University, Ames, Iowa 50011, USA}

\author{Shiya Chen}
\affiliation{Department of Physics, Xiamen University, Xiamen 361005, China}
\author{Weiyi Xia}
\affiliation{Department of Physics and Astronomy, Iowa State University, Ames, Iowa 50011, USA}
\affiliation{Ames Laboratory-USDOE, Iowa State University, Ames, Iowa 50011, USA}

\author{Wei-Shen Tee}
\affiliation{Department of Physics and Astronomy, Iowa State University, Ames, Iowa 50011, USA}
\affiliation{Ames Laboratory-USDOE, Iowa State University, Ames, Iowa 50011, USA}
\author{Yang Sun}
\affiliation{Department of Physics, Xiamen University, Xiamen 361005, China}
\author{Cai-Zhuang Wang}
\affiliation{Department of Physics and Astronomy, Iowa State University, Ames, Iowa 50011, USA}
\affiliation{Ames Laboratory-USDOE, Iowa State University, Ames, Iowa 50011, USA}

\author{Vladimir Antropov}
\affiliation{Department of Physics and Astronomy, Iowa State University, Ames, Iowa 50011, USA}
\affiliation{Ames Laboratory-USDOE, Iowa State University, Ames, Iowa 50011, USA}

% \affiliation{Department of Physics and Astronomy, Iowa State University, Ames, Iowa 50011, USA}
% \affiliation{Department of Physics, Xiamen University, Xiamen 361005, China}
% \affiliation{Ames Laboratory-USDOE, Iowa State University, Ames, Iowa 50011, USA}

\date{\today}% It is always \today, today,

\begin{abstract}

A wide variety of chemical substitutions in ferromagnetic Kagome systems can lead to diverse magnetic phases with electronic structures suitable for topological or quantum material properties. Here, we study the electronic structure and magnetic orderings using first-principles calculations for the magnetic Kagome compounds \ce{ZrFe6Ge6}, \ce{ZrFe6Ge4}, and \ce{ZrFe6Ge5}. For \ce{ZrFe6Ge6}, the obtained ground-state magnetic structure is A-type antiferromagnetic (AFM), in agreement with existing experiments. We predicted that the magnetic ground states of \ce{ZrFe6Ge4} and \ce{ZrFe6Ge5} are collinear A-type bilayer AFM structures with long-period ordering that involves a mix of FM and AFM interlayer orientations. The formation of such long-range magnetic structures appears to be a general feature and is not tied to specific substitutions. The magnetic moments in these systems are largely local and only weakly dependent on the magnetic configuration, with magnitudes in good agreement with available experimental estimates. Neutron scattering experiments, which could provide direct verification of these predictions, are therefore of particular importance.

\end{abstract}

\maketitle

%\tableofcontents

\section{Introduction}

Magnetic Kagome materials have attracted significant research interest as platforms for exploring the interplay among band topology, correlated electrons, and magnetism. Of these systems, the \ce{AT6X6} family of structures (where A is a cation, T is the Kagome transition metal, and X is an element in the p-block) exhibits a wide variety of topological features, such as Dirac points, van Hove singularities, flat bands, and magnetic states, ranging from ferromagnetism to antiferromagnetism to spin spirals \cite{wang2023quantum,yin2020quantum,venturini1993magnetic,chen2021large,el1991crystal,peng2021realizing,hu2022tunable}. Some \ce{AT6X6} structures have been observed to possess antiferromagnetism with Néel temperatures above room temperature \cite{venturini1992crystallographic,venturini1991magnetic,venturini1993magnetic,li2021dirac,mazet2013magnetic,mazet2000neutron,mulder1993155gd}.  In addition, these systems exhibit both easy-plane anisotropies and easy-axis anisotropies \cite{venturini1992crystallographic,venturini1991magnetic,el1994magnetic,ghimire2020competing}. 

Structure variants of the \ce{AT6X6} family that include a kagome structure also exist, namely \ce{AT6X5} and \ce{AT6X4}. These structures have a higher density of smagnetic atoms than \ce{AT6X6}, and some enhancement of magnetic properties is expected. Additionally, these structures are understudied compared to the \ce{AT6X6} structures. Among these systems, only \ce{ZrFe6Ge4}, \ce{ScFe6Ge4}, \ce{LiFe6Ge4}, and \ce{LiFe6Ge5} have demonstrated their magnetic properties experimentally \cite{matar2015chemical,kassem2024new,mantravadi2024experimental}, while their topological features have not been studied at all. Thus, theoretical exploration of the electronic structure of these potentially promising topological and spintronics systems is desirable.  

While experimental work remains limited, recent computational studies have laid the groundwork for characterizing the magnetic properties of these systems.  A high-throughput study \cite{chen2026computational} recently proposed many new \ce{AT6X6} systems that are stable across a variety of magnetic orderings. Other computational studies have also investigated the variant \ce{AT6X4} and \ce{AT6X5} structures \cite{zhang2025competing,vishina2025m,matar2015chemical,kassem2024new}.
In some cases, however, computational predictions and experimental measurements appear to be in conflict, with calculations predicting an FM ground state while the measured magnetization saturation is much lower than expected for a typical ferromagnet \cite{vishina2025m,matar2015chemical}. 

A possible explanation for this discrepancy is the presence of complex long-range interlayer magnetic interactions in these systems. The nearest-layer magnetic interaction, which was assumed in computational studies \cite{vishina2025m,matar2015chemical}, does not seem to be a good model for these systems, as complex magnetic coupling, including double-helix structures, is commonly observed in many \ce{AT6X6} systems \cite{venturini1992crystallographic,venturini1991magnetic,li2021dirac,mazet2013magnetic}. Recent works also suggested deviation from the nearest-layer model of the exchange interactions in \ce{AT6X6} \cite{chen2026computational,zhang2025competing}.
So, it seems logical that, for the \ce{AT6X5} and \ce{AT6X4} systems, the corresponding studies should also include more long-range magnetic orderings.

In this work, we use electronic structure calculations to determine the ground-state magnetic ordering and electronic structures of \ce{ZrFe6Ge6}, \ce{ZrFe6Ge5}, and \ce{ZrFe6Ge4}. It was found that each of these systems exhibits a different collinear AFM ground state. Within the plane, the Fe atoms exhibit a typical strong FM interaction, whereas the planes exhibit AFM or FM interactions depending on the stacking distance and the presence of intervening non-magnetic layers. This dependence on stacking distance and on intervening non-magnetic layers causes frustration in the exchange couplings between different layers, which in turn leads to long-period AFM ordering. In contrast to the earlier predicted  \cite{vishina2025m,matar2015chemical} FM ground state for \ce{ZrFe6Ge4}, we find that this system is stable in bilayer AFM. We find that the magnetic ground state of \ce{ZrFe6Ge6} corresponds to A-type AFM ordering, in agreement with previous experiments \cite{mazet2000neutron}, while \ce{ZrFe6Ge5} demonstrates bilayer AFM ordering similar to \ce{ZrFe6Ge4}.

\section{Methods}

We conducted spin-polarized density functional theory (DFT) calculations using the VASP package \cite{kresse1996efficient}, which employs the projector-augmented-wave (PAW) method. For the exchange-correlation energy, we used the Perdew-Burke-Ernzerhof (PBE) \cite{perdew1996generalized} generalized gradient approximation  (GGA). We used a plane-wave basis set with a kinetic energy cutoff of 600 eV. A Gaussian smearing of 0.05 eV was used. The convergence criteria were $10^{-5}$ eV for electronic self-consistence and $10^{-2}$ eV/\AA~ for ionic relaxation. A $\Gamma$-centered k-point mesh with a spacing of $2\pi \times 0.02$ \AA$^{-1}$ was used to sample the Brillouin zone in the structural optimization, total energy, and magnetic moment calculations. For the electronic density of states calculations, a k-spacing of $2\pi \times 0.01$ \AA$^{-1}$ was used. To capture long-period magnetic ordering between layers, we use a $1 \times 1 \times 2$ supercell of the conventional cell for all structures. To calculate the magnetic anisotropy of the identified lowest-energy magnetic configuration, we perform spin-polarized calculations with spin-orbit coupling turned on, with the spins aligned along the x, y, and z axes. 

\section{Results}

\subsection{Crystal Structure}

\ce{ZrFe6Ge6} is space group $P6/mmm$ and has a \ce{HfFe6Ge6}-type crystal structure \cite{mazet2000neutron}. That is, it consists of alternating layers of \ce{Fe3Ge3} and \ce{ZrGe2}.  \ce{Fe3Ge3} layers are arranged similarly to the structure of \ce{FeGe}, consisting of a honeycomb \ce{Ge2} lattice sandwiched between two vertically stacked \ce{FeGe2} layers. The \ce{Fe2Ge} layers consist of a kagome \ce{Fe} lattice with a \ce{Ge} atom occupying the center of the hexagons. The \ce{Ge2} lattice is arranged such that the \ce{Ge} atoms are centered above the \ce{Fe} triangles. The \ce{ZrGe2} layers consist of a \ce{Ge2} honeycomb lattice with a \ce{Zr} atom occupying the center of each hexagon. These layers are again stacked such that the Ge atoms lie above the center of the Fe triangles and the Zr atoms lie above the Ge atoms in the \ce{FeGe2} layers.

\ce{ZrFe6Ge4} has space group $R\overline{3}m$ and has the \ce{RFe6Ge4}-type crystal structure as outlined in Ref. \cite{matar2015chemical} and \cite{vishina2025m}.  It consists of an alternating pattern of \ce{FeGe2} bilayers and \ce{ZrGe2} layers. Within the bilayers, the \ce{FeGe2} layers are stacked with a $(2/3, 1/3)$ horizontal shift, such that the Ge atoms occupy the void above and below the Fe triangles of the other layer. The bilayers are stacked in an AB-BC-CA fashion, with - representing a \ce{ZrGe2} layer; that is, the bottom layer of the bilayers is aligned horizontally with the top layer of the previous bilayer. The \ce{ZrGe2} layer between them is stacked in the same fashion as in \ce{ZrFe6Ge6}.

As experimental information about \ce{ZrFe6Ge5} synthesis is unknown to us, we used a similar approach to that of Ref. \cite{chen2026computational} and identified \ce{ZrFe6Ge5} as a low-energy metastable phase lying 9 meV/atom above the convex hull. Notably, experimental databases such as ICSD \cite{zagorac2019recent} show that many successfully synthesized materials are metastable. For instance, the median energy above hull in observed metastable phases is 15 meV/atom in oxides and 67 meV/atom in nitrides \cite{sun2017thermodynamic}. Our energy above the convex hull is therefore below the median of known synthesized metastable compounds. In addition, hull energies are subject to a precision on the order of several meV \cite{horton2025accelerated}. On this order, small computational errors such as missing magnetic orderings, disorder, or correction scheme differences may change whether a phase is on or slightly above the hull. Thus \ce{ZrFe6Ge5} remains a strong synthesis candidate. 

\begin{figure*}[!ht]
\includegraphics[width=\linewidth]{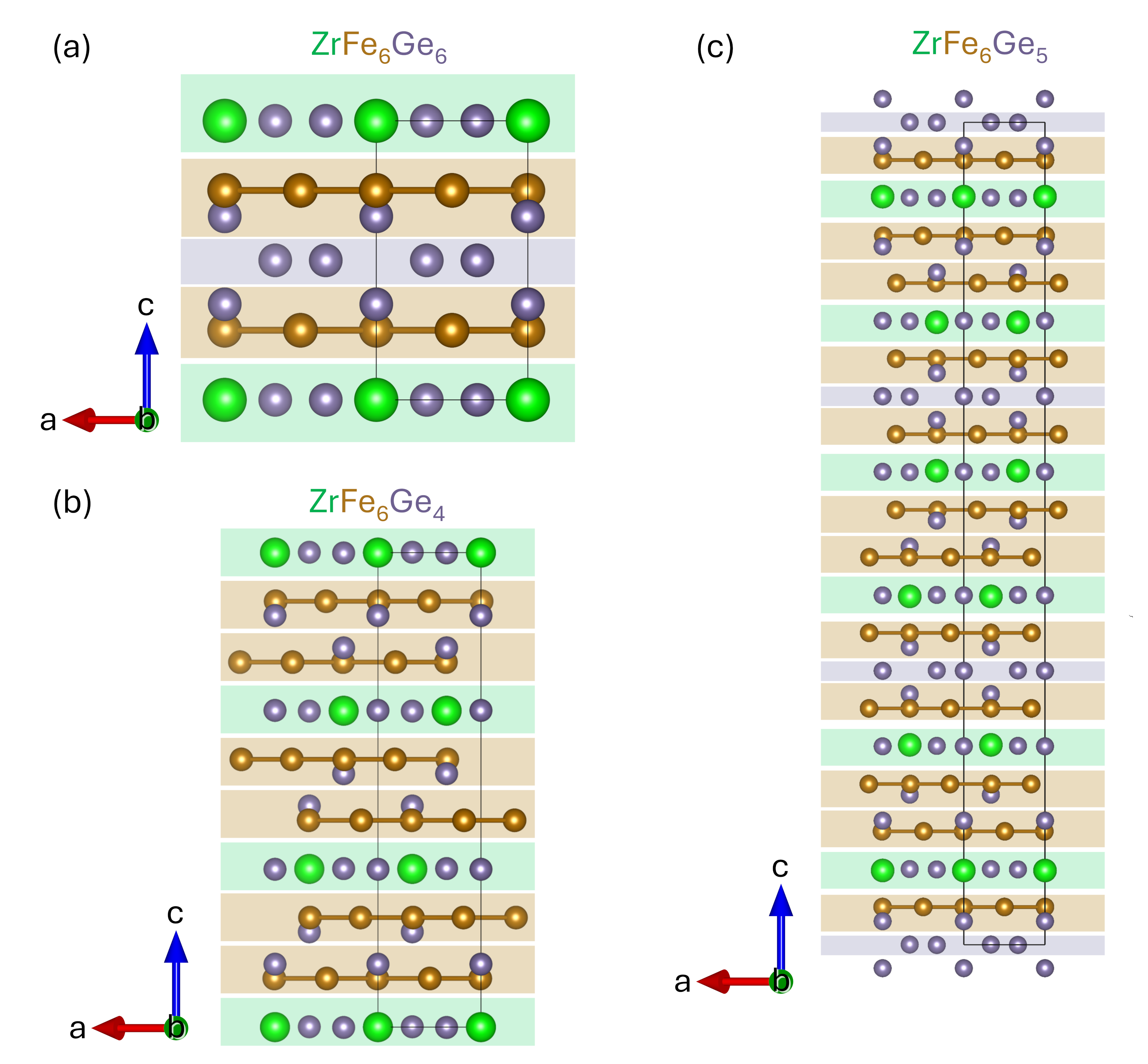}
\caption{\label{fig:structure} Crystal structures of (a) \ce{ZrFe6Ge6}, (b) \ce{ZrFe6Ge4}, and (c) \ce{ZrFe6Ge5} with layers indicated. Zr, Fe, and Ge atoms are colored green, orange, and purple, respectively. Light green, light orange, and light purple boxes mark \ce{ZrGe2}, \ce{Fe3Ge}, and \ce{Ge2} layers.}
\end{figure*}

The crystal structure of \ce{ZrFe6Ge5} adopts the same rhombohedral space group as \ce{ZrFe6Ge4}, which is closely related to the Li-Fe-Ge structure type \cite{mantravadi2024experimental,zhang2025competing}. The structure can be thought of as an alternating stack of the \ce{ZrFe6Ge6} and \ce{ZrFe6Ge4} crystal structures. That is, it follows a repeating pattern of \ce{Fe2Ge4 / ZrGe2 / Fe3Ge3 / ZrGe2} stacks.  The \ce{Fe3Ge3} layers are each stacked vertically above the top layer of the previous bilayer, and bilayers are always stacked directly vertically over the \ce{Fe3Ge3} layers. Thus, the lattice follows a CA-A0A-AB-B0B-BC-C0C stacking pattern, where -, as before, represents a \ce{ZrGe2} layer and 0 here represents a \ce{Ge2} layer. Note that the \ce{ZrGe2} and \ce{Ge2} layers are arranged in the same way in \ce{ZrFe6Ge6} and \ce{ZrFe6Ge5}. These structures are shown in Fig. \ref{fig:structure}.

These structures have two distinct types of interlayer distances. First, in the bilayers, the Fe atoms are separated by a distance close to 3\AA. Second,  across \ce{Ge2} layers, with or without a Zr atom, the Fe atoms are separated by a distance close to 4\AA.

\subsection{Collinear magnetic ground states}

\begin{figure*}[!ht]
\includegraphics[width=1.0\textwidth]{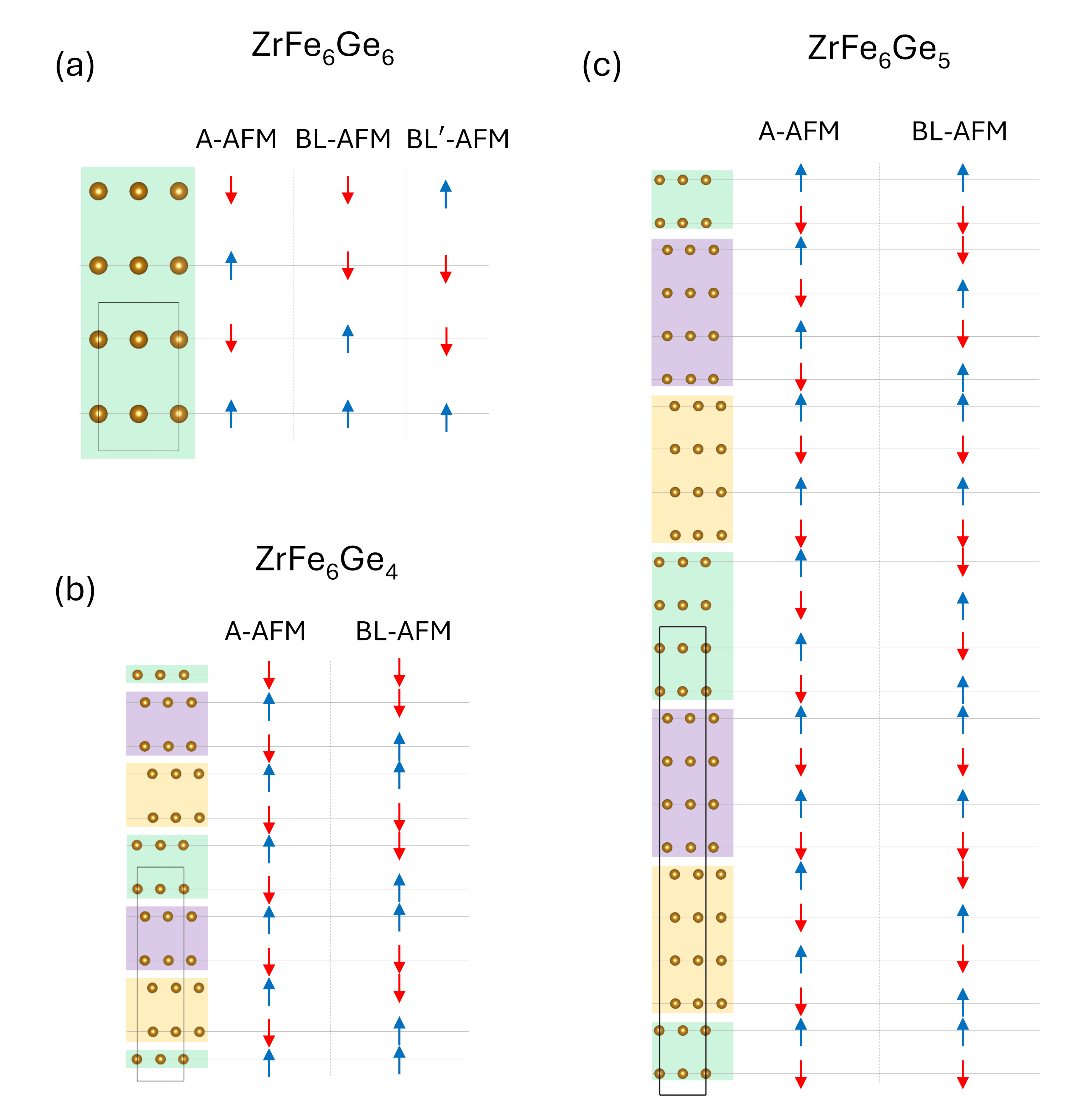}
\caption{\label{fig:ss} Magnetic structures for the AFM configurations of (a) \ce{ZrFe6Ge6}, (b) \ce{ZrFe6Ge4}, and (c) \ce{ZrFe6Ge5}. Fe layers are shown with other atoms omitted. Spin-up (down) layers are indicated by blue (red) arrows on the corresponding black line. Colored boxes represent blocks with the same lateral shift.}
\end{figure*}

\begin{table}[!h]
    \centering
    \begin{tabular}{|lcccc|}
\hline
\multicolumn{1}{|l|}{}                               & \multicolumn{1}{c|}{FM}         & \multicolumn{1}{c|}{A-AFM}         & \multicolumn{1}{c|}{BL-AFM}                               & BL$' $-AFM \\ \hline
\multicolumn{1}{|l|}{\ce{ZrFe6Ge6}} & \multicolumn{1}{c|}{39.38}      & \multicolumn{1}{c|}{\textbf{0.0}} & \multicolumn{1}{c|}{24.48}                                & 23.87   \\ \hline
\multicolumn{1}{|l|}{\ce{ZrFe6Ge5}} & \multicolumn{1}{c|}{37.89}      & \multicolumn{1}{c|}{10.53}         & \multicolumn{1}{c|}{\textbf{0.0}}       & --      \\ \hline
\multicolumn{1}{|l|}{\ce{ZrFe6Ge4}} & \multicolumn{1}{c|}{47.78}      & \multicolumn{1}{c|}{16.71}         & \multicolumn{1}{c|}{\textbf{0.0}}       & --      \\ \hline

\multicolumn{5}{|l|}{Moment}  \\ \hline
\multicolumn{1}{|l|}{\ce{ZrFe6Ge6}} & \multicolumn{1}{c|}{1.95}       & \multicolumn{1}{c|}{\textbf{2.04}} & \multicolumn{1}{c|}{2.02}                                 & 2.0    \\ \hline
\multicolumn{1}{|l|}{\ce{ZrFe6Ge5}} & \multicolumn{1}{c|}{1.95, 2.15} & \multicolumn{1}{c|}{1.99, 2.20}    & \multicolumn{1}{c|}{\textbf{2.02, 2.20}} & 
--      \\ \hline
\multicolumn{1}{|l|}{\ce{ZrFe6Ge4}} & \multicolumn{1}{c|}{2.20}       & \multicolumn{1}{c|}{2.19}          & \multicolumn{1}{c|}{\textbf{2.17}}       & --      \\ \hline

\end{tabular}
    \caption{Total energy (meV per Fe atom) above the magnetic ground state (indicated in bold)
    and local magnetic moments ($\mu_B$) of Fe sites for various magnetic configurations computed using GGA pseudopotentials. For \ce{ZrFe6Ge5}, two magnetic moments are indicated for the two inequivalent Fe sites.
    }
    \label{tab:deltaE}
\end{table}

To determine the collinear ground states, we perform self-consistent energy calculations for the FM and multiple AFM configurations in each phase. These are shown in Fig. \ref{fig:ss}.  For \ce{ZrFe6Ge6}, we consider AFM configurations where the spins of Fe layers are antiparallel across both \ce{ZrGe2} and \ce{Ge2} layers (A-AFM), where the spins are parallel across \ce{Ge2} layers and antiparallel across \ce{ZrGe2} layers (BL-AFM), and where the spins are antiparallel across \ce{Ge2} layers and parallel across \ce{ZrGe2} layers (BL$' $-AFM). For \ce{ZrFe6Ge4} and \ce{ZrFe6Ge5}, we consider AFM configurations where the spins of the bilayers are either arranged parallel (BL-AFM) or antiparallel to each other (A-AFM).

The total energies of each spin configuration are shown in Table \ref{tab:deltaE}. We find the collinear ground state to be A-AFM in \ce{ZrFe6Ge6} and BL-AFM in \ce{ZrFe6Ge4} and \ce{ZrFe6Ge5}. Notably, BL-AFM has a period of 4 magnetic layers in \ce{ZrFe6Ge4} and a period of 8 magnetic layers in \ce{ZrFe6Ge5}. In these configurations, the spins in the bilayers are aligned parallel, while across the \ce{ZrGe2} and \ce{Ge2} layers, the Fe layers are antiparallel. This strongly indicates FM coupling in the bilayers, whereas the coupling across the \ce{ZrGe2} and \ce{Ge2} layers is AFM. 

For the ground states, the magnetic moments of the Fe atoms are 2.04 $\mu_B$ for \ce{ZrFe6Ge6}, 2.17 $\mu_B$ for \ce{ZrFe6Ge4}, and 2.02 $\mu_B$ and 2.20 $\mu_B$ for \ce{ZrFe6Ge5}. Paired with their higher density of magnetic ions, their larger local magnetic moments indicate that \ce{ZrFe6Ge4} and \ce{ZrFe6Ge5} have a strictly larger sublattice magnetization than \ce{ZrFe6Ge6}. 

A breakdown of the magnetic moments on each unique Fe site for each magnetic state is also shown in Table \ref{tab:deltaE}. The magnitudes of the Fe sites' moments depend only weakly on the configuration, indicating that they behave as local Heisenberg-type moments rather than itinerant moments. In all cases, the local magnetic moments on Zr and Ge sites are small, with the largest being 0.117 $\mu_B$ on Zr for BL$'$-AFM in \ce{ZrFe6Ge6}, and -0.088 on Ge for FM and BL-AFM in \ce{ZrFe6Ge4}. The magnetic moments on these sites are itinerant, being induced by the local Fe moments. Zr ions receive a small positive induced moment, while Ge ions receive a negative induced moment. For configurations where the induced moments receive opposite contributions from symmetrically equivalent Fe atoms, these induced moments cancel out, while in other configurations the contributions add together to induce comparatively larger moments, as is the case with Zr for BL$'$-AFM in \ce{ZrFe6Ge6}.

\begin{table}[!h]
    \centering
    \begin{tabular}{|lcccc|}
\hline
\multicolumn{1}{|l|}{}                               & \multicolumn{1}{c|}{FM}         & \multicolumn{1}{c|}{A-AFM}         & \multicolumn{1}{c|}{BL-AFM}                               & BL$' $-AFM \\ \hline
\multicolumn{1}{|l|}{\ce{ZrFe6Ge6}} & \multicolumn{1}{c|}{29.61}      & \multicolumn{1}{c|}{\textbf{0.00}} & \multicolumn{1}{c|}{16.01}                                & 14.27   \\ \hline
\multicolumn{1}{|l|}{\ce{ZrFe6Ge5}} & \multicolumn{1}{c|}{44.39}      & \multicolumn{1}{c|}{23.68}         & \multicolumn{1}{c|}{\textbf{0.00}}       & --      \\ \hline
\multicolumn{1}{|l|}{\ce{ZrFe6Ge4}} & \multicolumn{1}{c|}{31.56}      & \multicolumn{1}{c|}{9.90}         & \multicolumn{1}{c|}{\textbf{0.00}}       & --      \\ \hline
\multicolumn{5}{|l|}{Moment}  \\ \hline
\multicolumn{1}{|l|}{\ce{ZrFe6Ge6}} & \multicolumn{1}{c|}{1.31}       & \multicolumn{1}{c|}{\textbf{1.49}} & \multicolumn{1}{c|}{1.37}                                 & 1.38    \\ \hline
\multicolumn{1}{|l|}{\ce{ZrFe6Ge5}} & \multicolumn{1}{c|}{1.31, 1.78} & \multicolumn{1}{c|}{1,48, 1.89}    & \multicolumn{1}{c|}{\textbf{1.53, 1.82}} & 
--      \\ \hline
\multicolumn{1}{|l|}{\ce{ZrFe6Ge4}} & \multicolumn{1}{c|}{1.74}       & \multicolumn{1}{c|}{1.86}          & \multicolumn{1}{c|}{\textbf{1.87}}       & --      \\ \hline
\end{tabular}
    \caption{Total energy (meV per Fe atom) above the magnetic ground state (indicated in bold)
    and local magnetic moments of Fe sites for various magnetic configurations computed using LDA pseudopotentials.
    }
    \label{tab:LDA}
\end{table}

Previous studies \cite{lin2020tunable} have shown that Fe-based FM Kagome systems can be described using traditional DFT approaches, such as with GGA or LDA pseudopotentials, whereas for Mn-based systems, the inclusion of Hubbard-type correlations is likely necessary \cite{zhang2020topological,sadhukhan2024topological}. Notably, neither LDA+U nor DMFT are able to properly describe the magnetic ground state of \ce{YMn6Sn6} \cite{sadhukhan2024topological}. Thus, we did not include LDA+U treatment in this study. We, however, performed self-consistent LDA calculations with a GGA-optimized lattice. In Table \ref{tab:LDA}, we show these results. While local moments on Fe atoms decrease along with the magnetic stabilization energies (as expected), our overall predictions of the ground states in these systems hold. 

For each identified magnetic ground state, we performed magnetic anisotropy calculations. We found that for each material, the uniaxial anisotropy is preferable (see Table \ref{tab:other}). The \ce{ZrFe6Ge_x} structures have large uniaxial anisotropies as compared to other \ce{AFe6Ge_x} structures \cite{zhang2025competing,vishina2025m}. This is likely because Zr is heavier than the cations in other \ce{AFe6Ge_x} structures, such as Li or Sc, and thus exhibits stronger spin-orbit coupling.  

Additionally, we estimated the relative strength of the spin-lattice coupling in AFM and FM cases. To do so, we compare the total energy of the ground AFM state when the structure is relaxed in an FM state to that when the structure is fully relaxed in the AFM state ($\Delta E_\mathrm{lat}$). These results are also shown in Table \ref{tab:other}. The magnitude of $\Delta E_\mathrm{lat}$ increases by a factor of three as Ge ions are removed from the compound, suggesting that \ce{ZrFe6Ge5} and \ce{ZrFe6Ge4} magneto-elastic coupling in these systems is strong and significantly modifies magnetic stabilization energy.

\begin{table}[!h]
    \begin{tabular}{|l|c|c|c|}\hline
         &  Ground State&  MAE & $\Delta E_\mathrm{lattice}$ \\ \hline
         \ce{ZrFe6Ge6}&  A-AFM&  0.21 (1.13)& 0.5 (2.69) \\ \hline
         \ce{ZrFe6Ge5}&  BL-AFM&  0.25 (1.51)& 1.47 (8.89)\\ \hline
 \ce{ZrFe6Ge4}& BL-AFM& 0.32 (2.07)& 1.58 (10.22) \\ \hline
    \end{tabular}
    \caption{Magnetic ground state, magnetic anisotropy energy, and relaxation energy difference between the FM and AFM states in meV/Fe ($\mathrm{MJ}/\mathrm{m}^3$).}
    \label{tab:other}
\end{table}

\subsection{Magnetic Exchange Parameters}

\begin{table}[!h]
    \begin{tabular}{|l|c|c|c|c|c|}\hline
         &  $J^{(1)}_\mathrm{ZrGe_2}$&  $J^{(1)}_\mathrm{Ge_2}$& $J^{(1)}_\mathrm{BL}$&$J^{(2)}_{\ce{BL}\text{-}\ce{ZrGe2}}$&$J^{(2)}_{\ce{Ge2}\text{-}\ce{ZrGe2}}$\\ \hline
         \ce{ZrFe6Ge6}&  -19.4&  -20.0& --&--&2.2\\ \hline
         \ce{ZrFe6Ge5}&  -26.6&  -16.1& 9.4&3.2&-2.2\\ \hline
 \ce{ZrFe6Ge4}& -39.1& --& 8.3&-4.1 &--\\ \hline
    \end{tabular}
    \caption{Heisenberg exchange coupling parameters between Fe ions and adjacent Fe layers (meV). Positive (negative) values correspond to FM (AFM) coupling.}
    \label{tab:Heisenberg-Model}
\end{table}

 To confirm the observed coupling behavior between magnetic layers, we calculated the interlayer magnetic exchange coupling parameters using total-energy calculations of FM and AFM states. This method is justified here as the amplitude of the atomic magnetic moments on Fe atoms is relatively preserved in different magnetic configurations (see discussion above). Specifically, we fitted an effective Heisenberg model of the form $-\frac{1}{2} \sum_{i,j} J_{i,j} m_i m_j$ to the different spin configuration energies, where $J_{i,j}$ is the exchange coupling and $m_i$ are unit spin vectors. We include exchange interactions between first- and second-nearest-neighbor magnetic layers. In this model, there are five unique layer-layer exchange parameters. The nearest neighbor couplings are $J^{(1)}_\mathrm{ZrGe_2}$, $J^{(1)}_\mathrm{Ge_2}$, and $J^{(1)}_\mathrm{BL}$, corresponding to adjacent Fe layers separated by a \ce{ZrGe2} layer, a \ce{Ge2} layer, or no intervening layers (i.e. within an Fe bilayer). The second-neighbor couplings are $J^{(2)}_{\ce{Ge2}\text{-}\ce{ZrGe2}}$ and  $J^{(2)}_{\ce{BL}\text{-}\ce{ZrGe2}}$, corresponding to Fe layers separated by \ce{Ge2}, \ce{Fe}, and \ce{ZrGe2} layers or \ce{Fe} and \ce{ZrGe2} layers. To fit these parameters, we calculated the energy of several additional magnetic configurations for \ce{ZrFe6Ge5} and \ce{ZrFe6Ge4} (see Supplemental Materials for details). The fitted exchange parameters are shown in Table \ref{tab:Heisenberg-Model}. As expected from the A-AFM ground state in \ce{ZrFe6Ge6} and the BL-AFM ground states in \ce{ZrFe6Ge5} and \ce{ZrFe6Ge4}, we find that $J^{(1)}_{\ce{ZrGe2}}$ and $J^{(1)}_{\ce{Ge2}}$ are consistently AFM while $J^{(1)}_{\ce{BL}}$ is consistently FM. While the strength of $J^{(1)}_{\ce{Ge2}}$ and $J^{(1)}_{\ce{BL}}$ is relatively consistent between structures, $J^{(1)}_{\ce{ZrGe2}}$ increases by a factor of two from \ce{ZrFe6Ge6} to \ce{ZrFe6Ge4}.
 
 In addition to interlayer exchanges, we attempted to determine the intralayer coupling between Fe ions by including magnetic configurations with AFM ordering within individual Fe layers. However, these configurations are highly unstable and consistently relaxed back to intralayer FM states. Thus, while the strength of the intralayer exchange cannot be determined by energy-mapping, we may conclude that it is strongly FM. 

\subsection{Electronic Structure}

\begin{figure*}[!ht]
\includegraphics[width=1.0\textwidth]{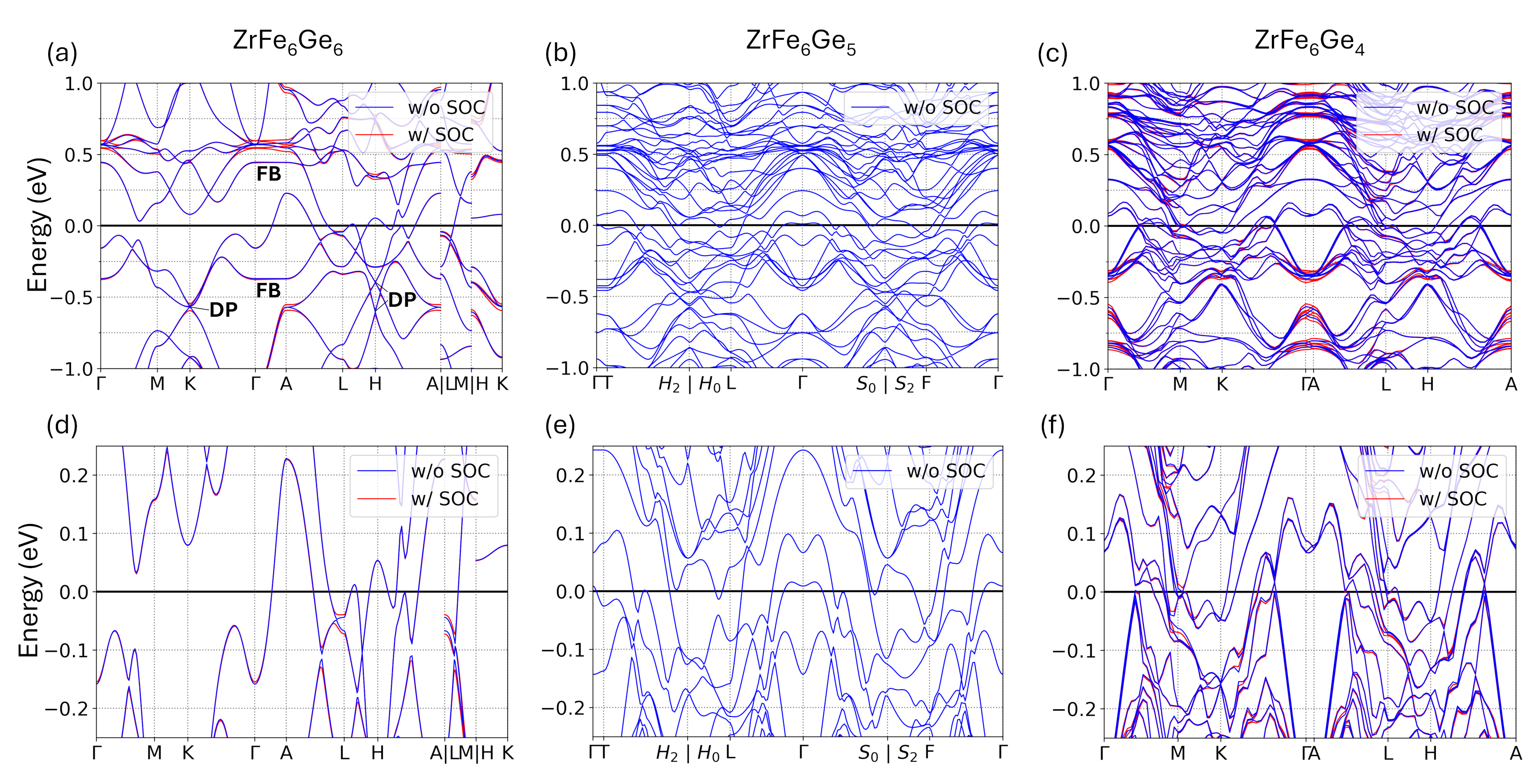}
\caption{\label{fig:bands} Spin-polarized electronic band structures of (a,d) \ce{ZrFe6Ge6}, (b,e) \ce{ZrFe6Ge5}, and (c,f) \ce{ZrFe6Ge4} near the Fermi level. Red and blue curves indicate bands calculated with and without spin-orbit coupling, respectively.}
\end{figure*}

\begin{figure*}[!ht]
\includegraphics[width=1.0\textwidth]{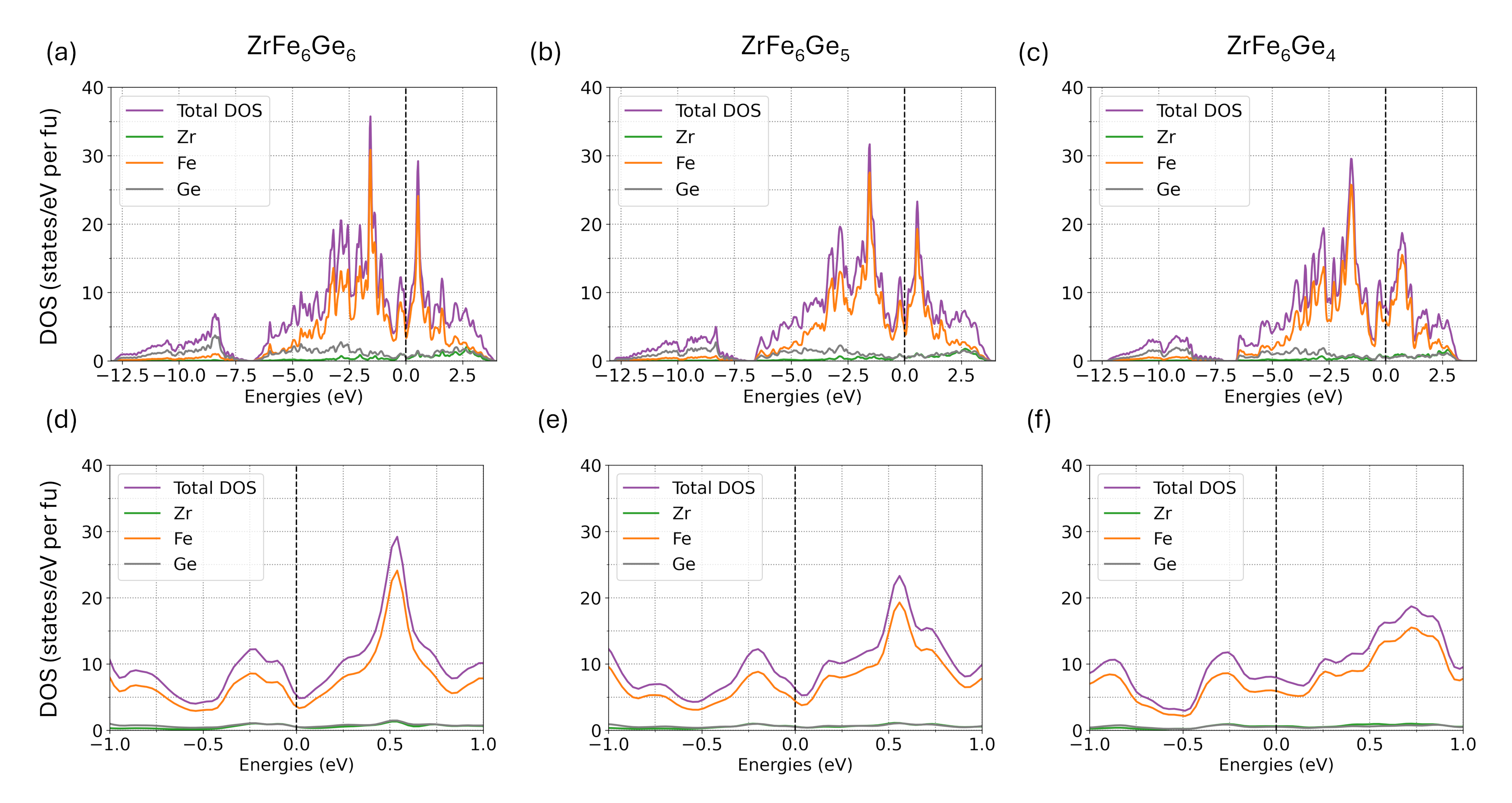}
\caption{\label{fig:dos} Total and element-projected DOS for (a,d) \ce{ZrFe6Ge6}, (b,e) \ce{ZrFe6Ge5}, and (c,f) \ce{ZrFe6Ge4}. (d-f) show the densities of states shown in (a-c) zoomed in near the Fermi level. Total DOS is shown in purple, and projected DOS for Ge, Fe, and Zr are shown in gray, orange, and green, respectively.}
\end{figure*}

The band structures for the collinear ground states of each system are shown in Figs. \ref{fig:bands}a-c. Near the Fermi level, both the relativistic and non-relativistic bands are displayed to show the effect of SOC. We find that all the systems are metallic. In \ce{ZrFe6Ge6}, flat bands are present at 0.37 eV below and 0.44 above the Fermi level along the $\Gamma$-A line. Dirac points are also present at the K point at 0.45 eV above and 0.57 eV below the Fermi level, and at the H point at both 0.4 and 0.6 eV below the Fermi level. These features are shown in Figs. \ref{fig:bands}d-e. Note that there is a Dirac-point-like structure that exists along the L-H line 0.1 eV below the Fermi level. However, the two bands repel each other, forming a sub-meV local gap that prevents the crossing required for a true Dirac point. Saddle points are present at M at 0.16 eV above the Fermi level and at L at 0.04 and 0.07 eV below the Fermi level. These flat-band features are, however, inaccessible via doping, requiring the removal of 3.6 electrons per formula unit to shift the Fermi energy into the lower flat band in the rigid-band approximation. Flat bands may be more accessible in a similar structure with fewer electrons, such as \ce{ZrFe6Ga6}. The band structures of \ce{ZrFe6Ge4} and \ce{ZrFe6Ge5} lack the characteristics of a Kagome lattice band structure.

The total and partial density of states are shown in Fig. \ref{fig:dos}. As the number of Ge ions in the structure decreases, we find that the system becomes more metallic, with the DOS at the Fermi level being 4.85 states/eV/f.u. for \ce{ZrFe6Ge6}, 6.03 states/eV/f.u. for \ce{ZrFe6Ge5}, and 7.96 states/eV/f.u. for \ce{ZrFe6Ge4}. For \ce{ZrFe6Ge6} there is a sharp peak in the Fe-atom projected DOS at 0.5 eV. As more Ge vacancies are added, this peak broadens and shifts upwards in energy, peaking at 0.75 eV for \ce{ZrFe6Ge4}. The broadening of this peak is likely due to the transition from Fe Kagome single layers to Fe Kagome bilayers. As more Fe bilayers are added to the system, interlayer interactions increase, causing the DOS peak to broaden. This can also be seen in the disappearance of the flat band at 0.5 eV along the $\Gamma$-A line when transitioning from \ce{ZrFe6Ge6} to \ce{ZrFe6Ge4}. As bilayers are added to the system, more vertical hopping becomes available, causing these states to no longer be constrained to a plane.

\section{Conclusion}

Here, we studied the electronic structure and magnetic properties of the magnetic Kagome compounds \ce{ZrFe6Ge6}, \ce{ZrFe6Ge4}, and \ce{ZrFe6Ge5}. \ce{ZrFe6Ge6} and \ce{ZrFe6Ge4} have previously been synthesized while, to our knowledge, \ce{ZrFe6Ge5} has not. However, we find \ce{ZrFe6Ge5} to be a strong synthesis candidate, given its very small metastability. Actual success, however, will still depend on kinetic accessibility, synthesis routes, and whether the convex hull is complete for the Zr-Fe-Ge chemical space. 

For \ce{ZrFe6Ge6}, we find the ground magnetic configuration to be A-type AFM, in agreement with past experiments. Although \ce{ZrFe6Ge6} exhibits some topological features in its band structure, these features are too far away from the Fermi level to be accessible by doping. For \ce{ZrFe6Ge4} and \ce{ZrFe6Ge5}, we predict A-type AFM configurations with antiferromagnetic ordering between FM bilayers. These configurations have periods of 4 and 8 magnetic layers, respectively, which are considerably longer than those previously considered for these systems. Surprisingly, these magnetic orderings appear not to be sensitive to the electron count, as identical magnetic orderings have been previously found in some Li- and Sc-based Kagome systems \cite{zhang2025competing,kassem2024new}. Thus, our studies suggest that the formation of long-range interactions between FM layers is a common feature of \ce{AT6X5} and \ce{AT6X4} FM Kagome structures. However, establishing the true range of the exchange interactions and their character can be a complicated task that would require full non-collinear analysis. Neutron scattering experiments are essential for the direct verification of our results. They may also clarify whether long-period ordering underlies the apparent discrepancy between prior computational studies and magnetization measurements.

\begin{acknowledgments}
Work at Ames National Laboratory was supported by the U.S. Department of Energy (DOE), Office of Science, Basic Energy Sciences, Materials Sciences and Engineering Division, including a grant of computer time at the National Energy Research Scientific Computing Center (NERSC), Berkeley, CA. Ames National Laboratory is operated for the U.S. DOE by Iowa State University under Contract No. DE-AC02-07CH11358.
\end{acknowledgments}

\appendix

%\nocite{*}

\bibliography{refs}% Produces the bibliography via BibTeX.

\end{document}

%\preprint{APS/123-QED}

\title{ Supplemental Materials: Antiferromagnetic Phases in Zr-Fe-Ge Kagome Systems}

\author{Peter Minch}
\affiliation{Department of Physics and Astronomy, Iowa State University, Ames, Iowa 50011, USA}
\affiliation{Ames Laboratory-USDOE, Iowa State University, Ames, Iowa 50011, USA}

\author{Shiya Chen}
\affiliation{Department of Physics, Xiamen University, Xiamen 361005, China}
\author{Weiyi Xia}
\affiliation{Department of Physics and Astronomy, Iowa State University, Ames, Iowa 50011, USA}
\affiliation{Ames Laboratory-USDOE, Iowa State University, Ames, Iowa 50011, USA}

\author{Wei-Shen Tee}
\affiliation{Department of Physics and Astronomy, Iowa State University, Ames, Iowa 50011, USA}
\affiliation{Ames Laboratory-USDOE, Iowa State University, Ames, Iowa 50011, USA}

\author{Yang Sun}
\affiliation{Department of Physics, Xiamen University, Xiamen 361005, China}
\author{Cai-Zhuang Wang}
\affiliation{Department of Physics and Astronomy, Iowa State University, Ames, Iowa 50011, USA}
\affiliation{Ames Laboratory-USDOE, Iowa State University, Ames, Iowa 50011, USA}

\author{Vladimir Antropov}
\affiliation{Department of Physics and Astronomy, Iowa State University, Ames, Iowa 50011, USA}
\affiliation{Ames Laboratory-USDOE, Iowa State University, Ames, Iowa 50011, USA}

% \affiliation{Department of Physics and Astronomy, Iowa State University, Ames, Iowa 50011, USA}
% \affiliation{Department of Physics, Xiamen University, Xiamen 361005, China}
% \affiliation{Ames Laboratory-USDOE, Iowa State University, Ames, Iowa 50011, USA}

\maketitle

\renewcommand{\thefigure}{S\arabic{figure}}
\renewcommand{\thetable}{S\arabic{table}}
\renewcommand{\theequation}{S\arabic{equation}}
\renewcommand{\thesection}{S\arabic{section}}

\setcounter{figure}{0}
\setcounter{table}{0}
\setcounter{equation}{0}
\setcounter{section}{0}

%\tableofcontents

\section{Additional Magnetic Configurations}

\begin{figure*}[!h]
\includegraphics[width=\linewidth]{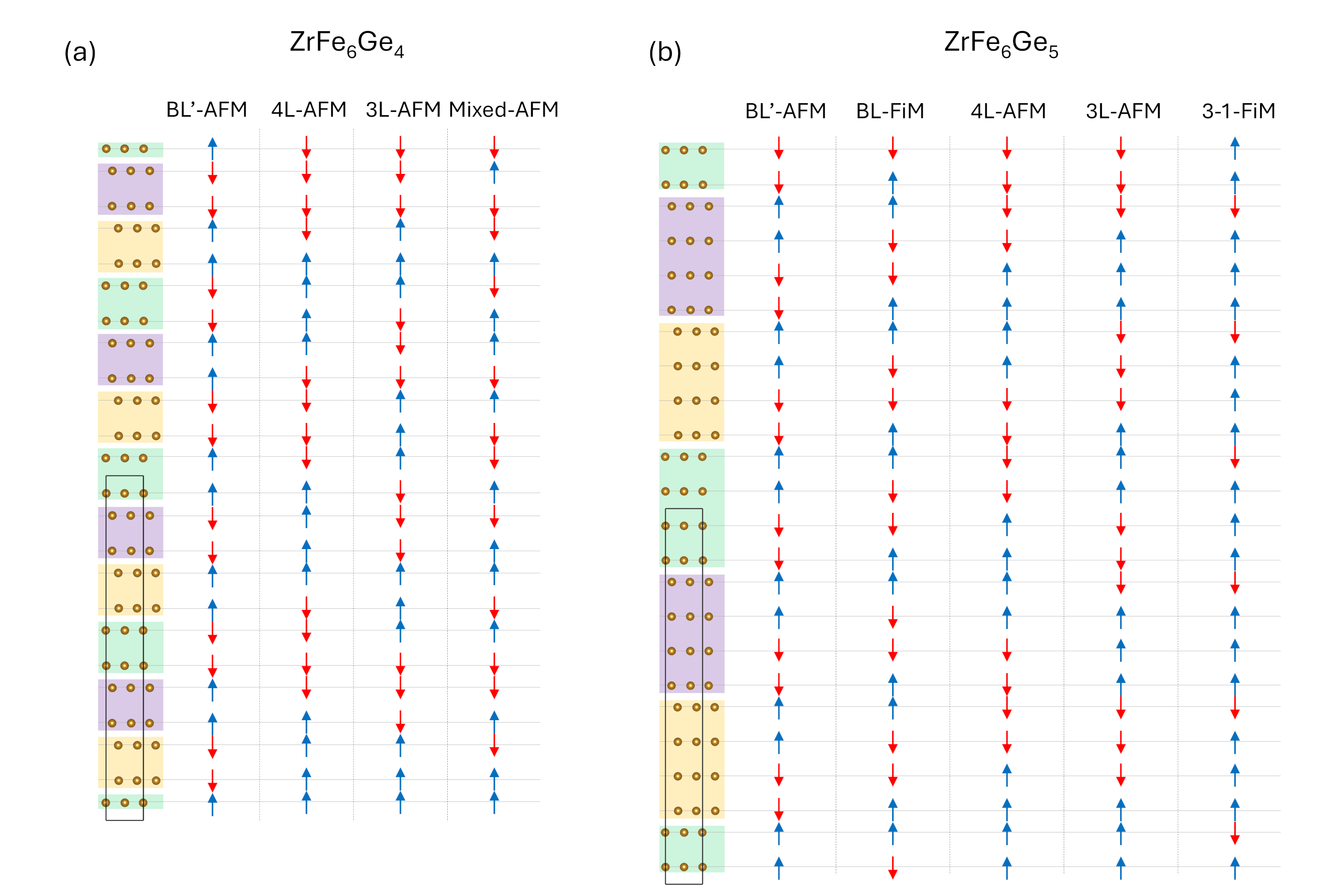}
\caption{\label{fig:supp-spin-states}Magnetic structures for additional AFM configurations of (a) \ce{ZrFe6Ge4} and (b) \ce{ZrFe6Ge5}. Fe layers are shown with other atoms omitted. Spin-up (down) layers are indicated with blue (red) on the corresponding black line. Colored boxes represent blocks with the same lateral shift.}
\end{figure*}

The additional antiferromagnetic (AFM) and ferrimagnetic (FiM) configurations for \ce{ZrFe6Ge4} and \ce{ZrFe6Ge5} that were calculated in order fit their Heisenberg models are shown in Fig. \ref{fig:supp-spin-states}. The Heisenberg model parameters were fit to the energies of the magnetic configurations using a least-squares fitting on a model of the form $-\frac{1}{2} \sum_{ij} J_{ij} S_i S_j$, where $J_{ij}$ are the exchange couplings and $S_i$ are unit spin vectors. For all magnetic configurations, their energies and the prefactors for each exchange coupling used in the fitting are shown in Table \ref{tab:HM-166}-\ref{tab:HM-164}.

\begin{table}[!h]
    \begin{tabular}{|l|c|c|c|c|}\hline
         \ce{ZrFe6Ge6}&   $\Delta E$&$J^{(1)}_\mathrm{ZrGe_2}$&  $J^{(1)}_\mathrm{Ge_2}$&$J^{(2)}_{\ce{Ge2}\text{-}\ce{ZrGe2}}$\\ \hline
         FM&   39.4&1/2&  1/2&1\\ \hline
         A-AFM&   0&-1/2&  -1/2&1\\ \hline
 BL-AFM&  24.5&-1/2& 1/2&-1\\ \hline
 BL'-AFM& 23.9& 1/2& -1/2&-1\\\hline
    \end{tabular}
    \caption{Total energy (meV per Fe atom) above the magnetic ground state for \ce{ZrFe6Ge6} and prefactors for each exchange term in the Heisenberg model.}
    \label{tab:HM-166}
\end{table}

\begin{table}[!h]
    \begin{tabular}{|l|c|c|c|c|c|c|}\hline
         \ce{ZrFe6Ge5}&   $\Delta E$&$J^{(1)}_\mathrm{ZrGe_2}$&  $J^{(1)}_\mathrm{Ge_2}$& $J^{(1)}_\mathrm{BL}$&$J^{(2)}_{\ce{BL}\text{-}\ce{ZrGe2}}$&$J^{(2)}_{\ce{Ge2}\text{-}\ce{ZrGe2}}$\\ \hline
         FM&   37.9&1/2&  1/4& 1/4&1/2&1/2\\ \hline
         A-AFM&   10.5&-1/2&  -1/4& -1/4&1/2&1/2\\ \hline
 BL-AFM&  0&-1/2& -1/4& 1/4&1/2&-1/2\\ \hline
 BL'-AFM& 35.4& 1/2& -1/4& -1/4& -1/2&-1/2\\\hline
 BL-FiM& 14.6& -1/2& 1/4& 1/4& -1/2&-1/2\\\hline
 4L-AFM& 31.5& 1/2& -1/4& 1/4& 1/2&-1/2\\\hline
 3L-AFM& 30.5& 1/6& 1/24& 1/24& -1/6&-1/6\\\hline
 3-1-FiM& 31.2& 0& 1/4& -1/4& 0&0\\\hline
    \end{tabular}
    \caption{Total energy (meV per Fe atom) above the magnetic ground state for \ce{ZrFe6Ge5} and prefactors for each exchange term in the Heisenberg model.}
    \label{tab:HM-165}
\end{table}

\begin{table}[!h]
    \begin{tabular}{|l|c|c|c|c|}\hline
         \ce{ZrFe6Ge4}&   $\Delta E$&$J^{(1)}_\mathrm{ZrGe_2}$& $J^{(1)}_\mathrm{BL}$&$J^{(2)}_{\ce{BL}\text{-}\ce{ZrGe2}}$\\ \hline
         FM&   47.8&1/2& 1/2&1\\ \hline
         A-AFM&   16.7&-1/2& -1/2&1\\ \hline
 BL-AFM&  0&-1/2& 1/2&-1\\ \hline
 BL'-AFM& 47.9& 1/2& -1/2& -1\\\hline
 4L-AFM& 24.5& 0& 1/2& 0\\\hline
 3L-AFM& 33.6& 1/6& 1/6& -1/3\\\hline
 Mixed-AFM& 10.4& -1/2& 0& 0\\\hline
    \end{tabular}
    \caption{Total energy (meV per Fe atom) above the magnetic ground state for \ce{ZrFe6Ge4} and prefactors for each exchange term in the Heisenberg model.}
    \label{tab:HM-164}
\end{table}

%\nocite{*}

\bibliography{refs}% Produces the bibliography via BibTeX.